\documentclass[fleqn,usenatbib]{mnras}

\usepackage[T1]{fontenc}
\usepackage{ae,aecompl}

\usepackage{graphicx}	
\usepackage{amsmath}	
\usepackage{amssymb}	

\usepackage{graphicx,color}	

\newcommand{\kms}{\textrm{km s}^{-1}}
\newcommand{\hi}{\text{H\,\sc{i}}}

\newcommand{\code}[1]{\texttt{#1}}

\title[Galaxy Asymmetries in 3D]{Measuring Galaxy Asymmetries in 3D}

\author[N. Deg et al.]{
N. Deg$^{1}$,\thanks{E-mail: nathan.deg@queensu.ca}
M. Perron-Cormier$^{1}$,
K. Spekkens$^{2,1}$,
M. Glowacki$^{3}$,
S.-L. Blyth$^{4}$,
N. Hank$^{5}$,
\\
$^{1}$Department of Physics, Engineering Physics, and Astronomy, Queen's University, Kingston, ON, K7L 3N6, Canada\\
$^{2}$Department of Physics and Space Science, Royal Military College of Canada, P.O.\ Box 17000, Station Forces Kingston ON K7K~7B4, Canada\\
$^{3}$International Centre for Radio Astronomy Research, Curtin University, Bentley, WA 6102, Australia\\
$^{4}$Department of Astronomy, University of Cape Town, Private Bag X3, Rondebosch 7701, South Africa\\
$^{5}$Kapteyn Astronomical Institute, University of Groningen, Landleven 12, 9747 AD Groningen, the Netherlands\\
}

\date{Accepted XXX. Received YYY; in original form ZZZ}

\pubyear{2020}

\begin{document}
\label{firstpage}
\pagerange{\pageref{firstpage}--\pageref{lastpage}}
\maketitle

\begin{abstract}
One of the commonly used non-parametric morphometric statistics for galaxy profiles and images \textcolor{black}{is the asymmetry statistic}.  With an eye to current and upcoming large \textcolor{black}{neutral hydrogen (\hi)} surveys, we develop a 3D version of the asymmetry statistic that can be applied to datacubes.  This statistic is more resilient to variations due to the observed geometry than \textcolor{black}{1D asymmetry measures}, and can be successfully applied to lower spatial resolutions \textcolor{black}{(3-4 beams \textcolor{black}{across the galaxy major axis})} \textcolor{black}{than} the 2D statistic.  We have also modified the asymmetry definition from an `absolute difference' version to a `squared difference' version that removes much of the bias due to noise contributions for low \textcolor{black}{signal-to-noise} observations.  Using a suite of mock asymmetric cubes we show that the background-corrected, squared difference \textcolor{black}{3D} asymmetry statistic can be applied to many marginally resolved galaxies in \textcolor{black}{large wide-area} \textcolor{black}{\hi} surveys \textcolor{black}{such as} WALLABY \textcolor{black}{on the Australian SKA Pathfinder (ASKAP).}
\end{abstract}




\section{Introduction}

One of the first ideas explored in extragalactic astronomy \textcolor{black}{was how to classify} galaxies based on their morphology. 
The most well-known are the Hubble schema \citep{Hubble1926} and the extended de Vaucouleurs system \citep{deVaucouleurs1959}, which classify galaxies into a few major classes; spirals, ellipticals \textcolor{black}{and lenticulars}, and irregulars.  \textcolor{black}{The Hubble sequence separates these classifications into early-types (ellipticals and lenticulars) and late-types (spirals, barred or not, and irregulars) based on ideas of galaxy evolution \textcolor{black}{at that time}.  While this early \textcolor{black}{association} has been shown to be broadly incorrect, the connection between visual appearance/classification and galaxy formation\textcolor{black}{/}evolution has continued to the present.  For instance,} the morphology of a galaxy has been found to \textcolor{black}{correlate with} the \textcolor{black}{gas content \citep{Roberts1994},} star formation rate \citep{Wuyts2011,Leslie2020}, star formation efficiency \citep{Saintonge2012,Ellison2018}, metallicity \citep{Ellison2008}, and more.  

As \textcolor{black}{observations} have become more sensitive and data volumes have expanded, many new galaxies \textcolor{black}{have been} discovered that defy simple classification into \textcolor{black}{early and late types}, leading to the number of irregulars \textcolor{black}{increasing} exponentially.  One approach, which has been used with great success, is to add new galaxy classes \textcolor{black}{(see \citet{Buta2013} for a review)}.  Regardless of the classification scheme devised, increased data volumes have made the visual classification of all observed galaxies very difficult.  At this point, no single person can visually classify all the galaxies detected in a single large survey.  One possible solution to this problem is the use of `citizen science' like the Galaxy Zoo project \citep{Willett2013,Willett2017}.  Alternatively, approaches involving machine learning have also shown a great deal of promise (for some examples see \citealt{Huertas-Company2015,Barchi2020,Walmsely2022} and references therein).

A different approach is to use \textcolor{black}{quantitative} non-parametric measurements to quantify galaxy morphologies.  One of the most successful approaches is the use of the CAS parameters (Concentration, Asymmetry, and Smoothness) pioneered by \textcolor{black}{\citet{Conselice2000} and  \cite{Conselice2003}}.  These are often coupled with the Gini and $M_{20}$ parameters of \citet{Lotz2004}.  The calculation of these measures can be automated rather simply, allowing them to be applied to large surveys with relative ease.  \textcolor{black}{For example,} the CAS parameters have been used to distinguish between early and late type galaxies \citep{Conselice2003,Lotz2004}.  \citet{Rodriguez-Gomez2019} compared mock images from the IllustrisTNG simulation \citep{Marinacci2018,Naiman2018,Nelson2018,Naiman2018,Springel2018} to those of the Pan-STARRS survey \citep{Chambers2016} and found that the overall morphologies of the \textcolor{black}{simulated galaxies match} observations, but there \textcolor{black}{are some disagreements} in the morphology-color and morphology-size relations.  
\textcolor{black}{\citet{Pearson2019} used these statistics to train machine learning algorithms to identify mergers and examine the effect of merging on the star formation rate. \citet{Pearson2022} applied this technique to the HSC-NEP survey (Hyper Subprime-Cam North Ecliptic Pole; \citealt{Goto2017,Oi2021}) to generate a merger catalogue for that field.}
Additionally, \citet{Bellhouse2022} explored the use of these parameters to analyze ram-pressure-stripping and post-starburst galaxies, 
as well as to investigate the connection between AGN activity, star formation, and the disturbances in the galaxies \citep{Zhao2022}.

The success of the asymmetry statistic at quantifying optical galaxy morphologies suggests that it might be equally successful at quantifying morphology of the \hi\ content of galaxies.  \hi\ generally extends further \textcolor{black}{from the galaxy centre} than the stellar disk \textcolor{black}{\citep{Koribalski2018}}, \textcolor{black}{making it} more susceptible to disturbances \textcolor{black}{such as} interactions and mergers \textcolor{black}{\citep{Bok2019,Deg2020}}, ram pressure stripping \textcolor{black}{\citep{GunnGott1972}}, tidal effects \textcolor{black}{\citep{Haynes1984}}, accretion \textcolor{black}{\citep{Sancisi2008}}, outflows \textcolor{black}{\citep{Fraternali2017}}.   \textcolor{black}{However, measuring the asymmetry statistic using \hi\ imaging has proven to be quite difficult.  The availability of large samples of resolved \hi\ images is scarce \textcolor{black}{and the dynamic range of these moment maps} \textcolor{black}{can be} \textcolor{black}{orders of magnitude} \textcolor{black}{lower than that of} \textcolor{black}{optical images.}  Moreover, the available \hi\ moment maps tend to have lower \textcolor{black}{signal-to-noise}, $S/N$, and angular resolution than optical imaging.   However, surveys like MIGHTEE-HI (the \hi\ emission project for the MeerKAT International GHz Tiered Extragalactic Exploration, \citealt{Jarvis2016,Maddox2021}), WALLABY (the Widefield ASKAP L-band Legacy All-sky Blind surveY, \citealt{Koribalski2020}), \textcolor{black}{and the WSRT-APERTIF imaging survey (the Westerbork Synthesis Radio Telescope - APERture Tile In Focus; \citealt{Adams2022})} will change this as they detect orders of magnitude more galaxies \textcolor{black}{that are spatially} resolved in \hi}.


One of the first attempts at \textcolor{black}{measuring} the asymmetry statistic, along with a set of other morphometric measures, in \hi\ data was by \citet{Holwerda2011}.  They examined THINGS observations (The \hi\ Nearby Galaxy Survey; \citealt{Walter2008}) and found that the asymmetry was particularly sensitive to disturbances \textcolor{black}{in} \hi\ disks, \textcolor{black}{and carried follow-up studies foccussing on other other surveys to further explore this phenomenon \citep{Holwerda2011b,Holwerda2011d,Holwerda2011e}.}  
More recently, \citet{Reynolds2021} applied the 2D asymmetry measurement to a sample of WALLABY pilot observations.  They identified a number of particularly asymmetric galaxies and examined them to determine if their disturbances were due to ram pressure stripping.  

\textcolor{black}{As discussed above, one} of the issues with calculating the asymmetry of \hi\ moment maps is the lower $S/N$ and \textcolor{black}{angular} resolution of radio observations \textcolor{black}{compared to optical imaging}.  In this regime, \textcolor{black}{applying} the correct background subtraction becomes increasingly important \citep{Giese2016,Reynolds2020,Thorp2021}.  To understand these issues, \citet{Giese2016} examined a suite of mock moment 0 images.  They found that the low $S/N$ of typical \hi\ observations introduced a bias in the asymmetry calculation and that the traditional background subtraction from \citet{Conselice2000} and \citet{Conselice2003} overcorrected the results.  As such, they developed an empirical \textcolor{black}{background} correction for the asymmetry.  Similarly, analyzing mock IFU observations from the Illustris \citep{Vogelsberger2014,Genel2014,Sijacki2015} and IllustrisTNG simulations, \citet{Thorp2021} developed an alternate background correction for stellar \textcolor{black}{mass} maps. \textcolor{black}{\citet{Bilimogga2022} used the mock \hi\ images and profiles constructed from the \textsc{EAGLE} simulations \citep{Schaye2015,Crain2015} to investigate the effect of noise on the measured asymmetry and found that relatively high $S/N$ and resolution were required for robust measurements (when the measurement has not been corrected for the noise).}

Another issue with calculating the asymmetry of \hi\ moment maps is the relative paucity of data.  \textcolor{black}{However, }there are orders of magnitude more \hi\ velocity profiles than there are \textcolor{black}{images of} \hi\ disks \textcolor{black}{due to large single dish surveys like HIPASS (HI Parkes All Sky Survey, \citealt{Barnes2001}) and ALFALFA (Arecibo Legacy FAST ALFA, \citealt{Griovanelli2005,Haynes2018}).}
As such, quantifying the asymmetry of \textcolor{black}{1D} velocity profiles has proven to be quite fruitful.  One of the first profile asymmetry statistics to be adopted is the `lopsidedness' measure proposed by \citet{Peterson1974}, which compares the ratio of the flux on the approaching and receding sides of a profile.  This statistic has been used in a variety of different studies that highlight the many drivers of \hi\ asymmetry.  For instance, \citet{Espada2011} found that a significant number of isolated galaxies are asymmetric, while \citet{Bok2019} found close pairs tend to be more asymmetric than isolated galaxies.  This is in contrast to \citet{Zuo2022} who found that massive merger galaxies have similar levels of asymmetry as their \textcolor{black}{non-merger} sample.  Thus, while mergers may drive asymmetry, there must be both a mass dependence and other drivers of asymmetry.  

\citet{Watts2020b} utilized lopsidedness to analyze the xGASS survey \textcolor{black}{(the extended GALEX Arecibo SDSS Survey, \citealt{Catinella2018})}.  They found that when they properly accounted for the effect of noise, 37\% of the galaxies detected \textcolor{black}{were} asymmetric.  Additionally, they found that satellite galaxies tended to be more asymmetric than central galaxies, indicating that environmental processes are a key driver of asymmetry.  These results were followed up by \citet{Watts2020} \textcolor{black}{who} explored the lopsidedness of velocity profiles constructed from the IllustrisTNG simulation.  They confirmed that, in the simulation, the satellite galaxy population tends to be more asymmetric than central galaxies.  While the excess asymmetry is driven by ram-pressure stripping in the satellites, the general drivers of asymmetry affect both populations of galaxies.  More recently, \citet{Watts2021} examined the lopsidedness of velocity profiles from the ALFALFA survey and the xGASS survey.  They found that asymmetric galaxies tend to be more gas-poor than symmetric galaxies with similar stellar masses.  This is only a small sampling of the increasingly large efforts aimed at using \textcolor{black}{1D} profile asymmetries to characterize galaxies.

Recently \citet{Deg2020} and \citet{Reynolds2020} developed a new `channel-by-channel' \textcolor{black}{1D} asymmetry statistic for velocity profiles that is \textcolor{black}{analogous} to the standard 2D asymmetry statistic.  While the lopsidedness/flux ratio measure is an integral quantity, this new measure is sensitive to more local disturbances.  Moreover, its similarity to the 2D asymmetry statistic allows for easier comparison of 1D and 2D measurements from simulations and observations.  \citet{Deg2020} found that this statistic provided a better agreement to visual classifications of asymmetric profiles than the lopsidedness statistic.  \citet{Reynolds2020} applied this statistic, along with other asymmetry measures, to a sample of galaxies from the LVHIS \textcolor{black}{(Local Volume \hi\ Survey, \citealt{Koribalski2018})}, VIVA \textcolor{black}{(VLA Imaging of Virgo Spirals in Atomic Gas, \citealt{Chung2009})}, and HALOGAS \textcolor{black}{(The Westerbork Hydrogen Accretion in LOcal GAlaxieS, \citealt{Heald2011})} surveys, and found that the measured asymmetry does depend on the environment, but did not find a strong trend with \hi\ mass.  More recently \citet{Glowacki2022} explored the use of this statistic in the SIMBA cosmological simulation \citep{Dave2019} and found the \hi\ mass has the strongest correlation with the profile asymmetry.  When the \hi\ mass is controlled, highly asymmetric galaxies were found to be more gas poor and have larger specific star formation rates than their symmetric counterparts.

The 2D and 1D `channel-by-channel' asymmetry statistics are quite similar to each other and have been used to analyze \hi\ observations from a variety of different surveys as well as simulations.  However, modern \hi\ surveys are generally interferometric in nature, and the most common data product is a 3D datacube {that \textcolor{black}{contains} both spatial and spectral information simultaneously}.  Given this, it is worthwhile to extend the asymmetry statistic to 3D data cubelets (a data cube containing only a single galaxy detection).  Cubelets contain both \textcolor{black}{morphological} and kinematic information, which allows for a larger variety of disturbances to be detected in a single measurement than either a \textcolor{black}{2D} moment 0 map or \textcolor{black}{a 1D} velocity profile.  Moreover, the noise of a cubelet tends to be uniform, which is simpler to account for than the non-uniform noise structure of its derived data products. \textcolor{black}{In addition}, \textcolor{black}{1D and 2D asymmetries tend to be unreliable in the low resolution, low $S/N$ regime that comprises most detections from wide-field \hi\ surveys \citep[e.g.][]{Giese2016, Reynolds2020, Bilimogga2022}.  As we will show in this paper, moving to 3D allows the asymmetry statistic to be applied to lower resolution and $S/N$ detections.}

In this paper we introduce the 3D asymmetry statistic.  In addition, we show that a switch from an `absolute difference' asymmetry to a `squared difference' asymmetry allows the effect of noise on a measurement to be quantified and accounted for in a significantly more rigorous fashion.  In Section \ref{Sec:Asymmetries} \textcolor{black}{we derive 3D asymmetry measures using both absolute differences and our preferred squared difference formalism.}  Section \ref{Sec:Geometry} explores how the 1D, 2D, and 3D statistics depend on the observed geometry of a galaxy as well as the resolution \textcolor{black}{of an observation}.  Section \ref{Sec:Noise} shows the effect of noise on asymmetry measures, while Section \ref{Sec:WallabyExample} applies the \text{3D asymmetry} statistic to a mock WALLABY-like sample.  Finally Section \ref{Sec:Conclusion} provides a discussion of these measures and our conclusions.

\section{Asymmetry statistics}\label{Sec:Asymmetries}

The \textcolor{black}{2D} asymmetry statistic was initially designed to be applied to optical images \citep{Schade1995,Conselice2000,Conselice2003} \textcolor{black}{but it} \textcolor{black}{can} be applied to any two dimensional density or flux map\textcolor{black}{; it has also been modified for 1D velocity profile analysis \citep[][]{Deg2020,Reynolds2020}.}
This section will first review both the 2D and 1D asymmetry definitions.  It will then describe the 3D asymmetry \textcolor{black}{using the standard `absolute difference' definition,} as well as a new `squared difference' definition that can account for noise in a more rigorous \textcolor{black}{fashion}.

\subsection{2D}\label{subsec:2DAsym}

The 2D asymmetry has been defined in a few different ways.  The most common is \textcolor{black}{from \citet{Conselice2000}}:
\begin{equation}\label{Eq:Basic2DEquation}
    A_{2D, abs}=\frac{\sum_{j,k}\left|f_{j,k}-f_{-j,-k} \right|}{\sum_{j,k} |f_{j,k}+f_{-j,-k}|}~,
\end{equation}
where $(j,k)$ are the pixel indices relative to a center of rotation and $f_{j,k}$ is the flux of the pixel $(j,k)$, and the summation is done over all pixels in \textcolor{black}{a masked region of an image}.  Effectively, Eq.~\ref{Eq:Basic2DEquation} computes a pixel-by-pixel normalized difference between an image and that same image rotated by 180$^{\circ}$.  
\textcolor{black}{An alternate definition used in \citet{Conselice2003} and more recently in \citet{Abruzzo2018} drops the absolute sign in the denominator:}
\begin{equation}\label{Eq:Basic2DEquation_NoAbs}
    A_{2D, abs}=\frac{\sum_{j,k}\left|f_{j,k}-f_{-j,-k} \right|}{\sum_{j,k} ( f_{j,k}+f_{-j,-k})}~,
\end{equation}
\textcolor{black}{The advantage of Eq. \ref{Eq:Basic2DEquation_NoAbs} is that it allows for a slightly simpler calculation of the effect of noise (although it is still difficult as discussed in Sec. \ref{subSec:AsymNoise} and Sec. \ref{Sec:Noise}).  For the remainder of this paper we will utilize Eq. \ref{Eq:Basic2DEquation_NoAbs} rather than Eq. \ref{Eq:Basic2DEquation} \textcolor{black}{when computing `absolute difference' asymmetries}.}

The 2D asymmetry was originally designed for optical observations where using a pixel as the rotation point is reasonable \textcolor{black}{(see the discussion on centering in \citealt{Conselice2000})}.  However, it can be generalized to use arbitrary coordinates provided that some method of interpolation is \textcolor{black}{applied to} the image.  \textcolor{black}{In \citet{Conselice2000} (and many other implementations), the center point is found by minimizing the asymmetry.  This is a relatively straightforward approach, but the point that minimizes the asymmetry does not necessarily correspond to a physically meaningful \textcolor{black}{location} like the dynamical center of the galaxy}\textcolor{black}{, which can be estimated from the 3D datasets that are the focus of this work (e.g. \citealt{Deg2022,Westmeier2022}).}  


\subsection{1D and 3D Asymmetries}\label{subsec:1DAsym}

 Eq.~\ref{Eq:Basic2DEquation} can be generalized as
\begin{equation}\label{Eq:GeneralAsymmetry}
    A_{abs}=\frac{P_{abs}}{Q_{abs}}~,
\end{equation}
where 
\begin{equation}
 P_{abs}=\sum_{i}^{N}\left|f_{i}-f_{-i} \right|   ~,
\end{equation}
and
\begin{equation}\label{Eq:Q_Abs}
    Q_{abs}=\sum_{i}^{N}\left(f_{i} +f_{-i}\right)~.
\end{equation}
In this notation, the 2D sum over $(j,k)$ pixels in Eq.~\ref{Eq:Basic2DEquation} is replaced with a generalized sum over all \textit{$N$ pixel pairs} across one or more dimensions, with $i$ representing the pixel index; in 2D, $f_{i}=f_{j,k}$ and $f_{-i}=f_{-j,-k}$. 

The idea of pair indexing rather than pixel indexing reveals the 1D asymmetry statistic clearly.  Rather than using pixel indices as in Eq. \ref{Eq:Basic2DEquation}, the profile asymmetry of \citet{Deg2020} and \citet{Reynolds2020} simply uses pairs of velocity channels.  The only difference between \textcolor{black}{$A_{2D,abs}$} and \textcolor{black}{$A_{1D,abs}$} mathematically is the mapping of a flux pair, $i$, to a pair of velocity channels rather than to a pair of pixels. Instead of pairing pixels around a particular rotation point, the 1D asymmetry pairs channels matched across some reference velocity.


Given this notation, moving to 3D is relatively straightforward.  Rather than a pixel (2D) or velocity (1D), the rotation point is some cell inside \textcolor{black}{the 3D cubelet} and the $i$'th flux pair maps to two locations that are $180^\circ$ apart spatially and equally distant from the reference velocity.  

\textcolor{black}{Nonetheless, working in 3D introduces some additional complications.} In particular, the construction of a mask is significantly more complex.  Related to this issue is the symmetry of the mask itself.  
\textcolor{black}{In} 3D, masks are usually constructed using complex algorithms that rely upon $S/N$ levels.  For instance, \textsc{SoFiA-2} \citep{Westmeier2021} is a commonly used tool for detecting extragalactic \hi\ sources and it constructs masks containing each source.  These masks are rarely symmetric, which poses a problem for asymmetry calculations as an asymmetric mask itself will affect the value of the asymmetry measured \textcolor{black}{(see Sec. \ref{subsec:Uncertainty} for an example)}.  For this reason, we recommend that any 3D mask be symmeterized about the chosen rotation point.  \textcolor{black}{This is trivial when using a specific rotation point.  However this symmetrization step will significantly slow down approaches that attempt to find the point that minimize the asymmetry (e.g. \citealt{Conselice2000,Deg2020}), as the mask will need to be recalculated about each trial rotation point.}  \textcolor{black}{During preliminary tests with our asymmetry implementation (see Sec. \ref{Sec:Implementation}), we found that re-symmeterizing the mask when minimizing the asymmetry led to $\approx 10$ times longer runtimes}.  

Before moving to a discussion of the noise, it is worth noting one other advantage of the notation shown in Eqs. \ref{Eq:GeneralAsymmetry}-\ref{Eq:Q_Abs}.  A physically meaningful rotation point, like the dynamical center, does not need to lie at a particular pixel/channel/cell.  Rather than considering integer pairs of pixels, it is possible to interpolate to arbitrary points within a pixel/channel/cell.  Following this, the pairs are simply defined as $x_{i}-x_{cent}=x_{cent}-x_{-i}$, $y_{i}-y_{cent}=y_{cent}-y_{-i}$, and, as mentioned above, $v_{i}-v_{sys}=v_{sys}-v_{-i}$ for the appropriate dimensions, \textcolor{black}{where ($x_{cent},y_{cent},v_{sys})$ is the chosen rotation point}.

\subsection{Dealing with noise}\label{subSec:AsymNoise}

The entire discussion thus far has defined asymmetries in 1, 2, and 3 dimensions for noiseless data.  
\textcolor{black}{In addition to introducing uncertainty (see Sec. \ref{subsec:Uncertainty}), noise also causes a bias in the measured asymmetry.}
When the data are noisy, the observed flux of some channel/pixel/cell \textcolor{black}{can be written as} $F_{i}=f_{i}+g_{i}$, where $f_i$ is the signal as above, and $g_{i}$ is the contribution of the noise to that pixel.  

In this case the measured asymmetry, $C_{m}$, becomes
\begin{equation}
    C_{m}=\frac{P_{m}}{Q_{m}}~,
\end{equation}
where
\begin{equation}\label{Eq:P_m_abs}
   P_{m}= \sum_{i}^{N}\left|f_{i}-f_{-i} +g_{i}-g_{-i} \right|~,
\end{equation}
and 
\begin{equation}\label{Eq:Q_m_noise}
   Q_{m}= \sum_{i}^{N} ( f_{i}+f_{-i} +g_{i}+g_{-i} ) ~.
\end{equation}
\textcolor{black}{If the noise is uniform, \textcolor{black}{then} $g_{i}$ can be treated as a random draw from a distribution with a mean of zero}.
For a sufficiently large number of pairs and low levels of noise (regardless of the precise noise distribution), \textcolor{black}{Eq. \ref{Eq:Q_m_noise} reduces to Eq. \ref{Eq:Q_Abs}}.  Unfortunately the effect of the noise cannot be easily separated out in Eq. \ref{Eq:P_m_abs} due to the \textcolor{black}{non-linearity of the absolute value function}.  The typical approach to account for this is to approximate the noise-corrected asymmetry as
\begin{equation}\label{Eq:AbsAsym_Background}
    A_{m}\approx C_m -\frac{B_{abs}}{Q_{m}}~,
\end{equation}
where 
\begin{equation}
    B_{abs}=\sum_{i}^{N} \left| g_{i}-g_{-i}\right|~.
\end{equation}
\textcolor{black}{It is reasonable to adopt a Gaussian noise distribution for well-calibrated interferometric radio observations,} so $g_{i}$ is a random draw from a Gaussian with a mean of zero and \textcolor{black}{a standard deviation} of $\sigma$. For Gaussian noise, $B_{abs}$ can be simplified to
\begin{equation}\label{Eq:AbsBackground}
    B_{abs}\approx \frac{2 \sigma N}{\sqrt{\pi}}~.
\end{equation}

As pointed out in \citet{Giese2016} and \citet{Thorp2021}, this approach, while quite successful at high $S/N$, results in an over-subtraction at lower $S/N$ observations. \citet{Thorp2021} noted that, due to the rules of modular subtraction $|R|-|S|\le |R-S|$, this over-subtraction is expected, but it can be quite severe.  There are numerous methods that have been developed to deal with this bias or to determine \textcolor{black}{the $S/N$ at which} this bias becomes important \citep{Giese2016,Watts2020b,Reynolds2020,Thorp2021}.

An alternate approach is to redefine asymmetry using the \textit{squared difference} of flux pairs rather than the \textit{absolute difference} \textcolor{black}{in Eq.~\ref{Eq:GeneralAsymmetry}}. \textcolor{black}{This is very similar to the `rms' asymmetry introduced in \citet{Conselice2000}}.  In the absence of noise we can rewrite the asymmetry equation as
\begin{equation}
    A_{sq}^{2}=\frac{P_{sq}}{Q_{sq}}~,
\end{equation}
where 
\begin{equation}
    P_{sq}=\sum_{i}^{N}\left(f_{i}-f_{-i} \right)^{2}
\end{equation}
and
\begin{equation}
    Q_{sq}=\sum_{i}^{N}\left(f_{i}+f_{-i} \right)^{2}
\end{equation}
The \textcolor{black}{adjustment of the asymmetry equation denominator to $Q_{sq}$ } is necessary to keep \textcolor{black}{$A_{sq}$} unitless and independent of the number of pairs.

For noisy data, $P_{sq,m}$ becomes
\begin{equation}
    P_{sq,m}=\sum_{i}^{N}\left(f_{i}-f_{-i} +g_{i}-g_{-i}\right)^{2}~,
\end{equation}
\textcolor{black}{and the measured denominator, $Q_{sq,m}$, becomes}
\begin{equation}
    Q_{sq,m}=\sum_{i}^{N}\left(f_{i}+f_{-i} +g_{i}+g_{-i}\right)^{2}~.
\end{equation}
Expanding and rearranging $P_{sq,m}$ slightly gives
\begin{multline}\label{Eq:PExpand}
     P_{sq,m}=\left[\sum_{i}^{N}\left(f_{i}-f_{-i}\right)^{2}\right] + \left[\sum_{i}^{N}\left(g_{i}-g_{-i}\right)^{2}\right]
    \\+2\left[\sum_{i}^{N}\left(f_{i}-f_{-i}\right)\left(g_{i}-g_{-i}\right)\right].   
\end{multline}
The \textcolor{black}{first term in the equation above is} simply $P_{sq}$, \textcolor{black}{while the third} goes to zero for sufficiently large $N$ and small $\sigma$.  The second \textcolor{black}{term}, $B_{sq}$, is \textcolor{black}{the contribution of the noise to $P_{sq,m}$ and can be approximated as}
\begin{equation}\label{Eq:SqDBackground}
 B_{sq}= \left< \sum_{i}^{N}\left(g_{i}-g_{-i}\right)^{2}\right> \approx 2 N \sigma^{2}~,
\end{equation}
for Gaussian noise \textcolor{black}{and sufficiently large N}.

In a similar manner, $Q_{sq,m}$ can be expanded and rearranged.  Since 
\begin{equation}
    \left<\sum_{i}^{N}\left(g_{i}-g_{-i}\right)^{2}\right]=\left[\sum_{i}^{N}\left(g_{i}+g_{-i}\right)^{2}\right>~,
\end{equation}
for Gaussian noise \textcolor{black}{and large N}, we find that
\begin{equation}\label{Eq:Asq}
    A_{sq}=\left(\frac{P_{sq,m}-B_{sq}}{Q_{sq,m}-B_{sq}}\right)^{1/2}~.
\end{equation}
\textcolor{black}{The $B_{sq}$ terms should be understood as the \textit{systematic contribution} of the noise to the asymmetry measurement, and subtracting them from the numerator and denominator removes this bias.  This is not the same as the \textit{random uncertainty} of the asymmetry measurement, which is also produced by noise.  We discuss random uncertainties in Sec. \ref{subsec:Uncertainty}}.

\textcolor{black}{It is again worth noting here that our `squared difference' method is similar to the `rms' method of \citet{Conselice2000}.  In that work, they found an improved correlation with galaxy color.  However, in the low $S/N$ regime of many \hi\ cubelets, we \textcolor{black}{find the} cleaner background correction of the squared difference method to be a great advantage.} \textcolor{black}{We compare the performance of `absolute difference' and `squared difference' 3D asymmetries in Section~\ref{Sec:Noise}.}

\section{3D Asymmetries and Noiseless Data}\label{Sec:Geometry}

As an observed quantity, the asymmetry statistic is subject to a host of observational effects/biases, many of which are caused by the observed geometry and resolution \citep{Giese2016,Deg2020}.  These effects are present regardless of the `intrinsic' asymmetry or the observed noise.  As such, it is important to understand how the 3D asymmetry statistic depends on the \textcolor{black}{asymmetry} viewing angle (relative to the galaxy orientation), \textcolor{black}{the disk} inclination, and \textcolor{black}{the resolution of the observation in the absence of noise. We explore this performance here, and then add noise in Section~\ref{Sec:Noise}}.  

In this section, we compare the 3D asymmetry to the 2D and 1D asymmetry using mock cubelets \textcolor{black}{with} a variety of different geometries and resolutions.  The mock cubes are generated using a modified version of the \textsc{MCGSuite} code\footnote{\href{https://github.com/CIRADA-Tools/MCGSuite}{https://github.com/CIRADA-Tools/MCGSuite}} (\citealt{Lewis2019}, Spekkens et al. in prep), which generates realistic mock \hi\ cubelets \textcolor{black}{of flat axisymmetric \hi\ disks} using \textcolor{black}{empirical} scaling relations.  \textcolor{black}{The key parameters for \textsc{MCGSuite} are the \hi\ mass, $M_{\hi}$, (which generates the rotation curve and surface density profile) and the  diameter, $D_{\hi}$, \textcolor{black}{measured in angular resolution elements which we henceforth call `beams'}.  The diameter is defined as twice the radius \textcolor{black}{in the plane of the disk}, $R_{\hi}$, where the unconvolved surface density equals 1 M$_\odot$ pc$^{-2}$.  $D_{\hi}$ is defined in beams as \textsc{MCGSuite} calculates the distance to the object such that $2 R_{\hi}$ in kpc subtends an angle equal to the target size in beams. In addition, the observed inclination and position angle of the disk, $i$, and $\phi$ respectively,  are \textsc{MCGSuite} \textcolor{black}{input} parameters.} 

\begin{figure*}
    \centering
    \includegraphics[width=0.9\textwidth]{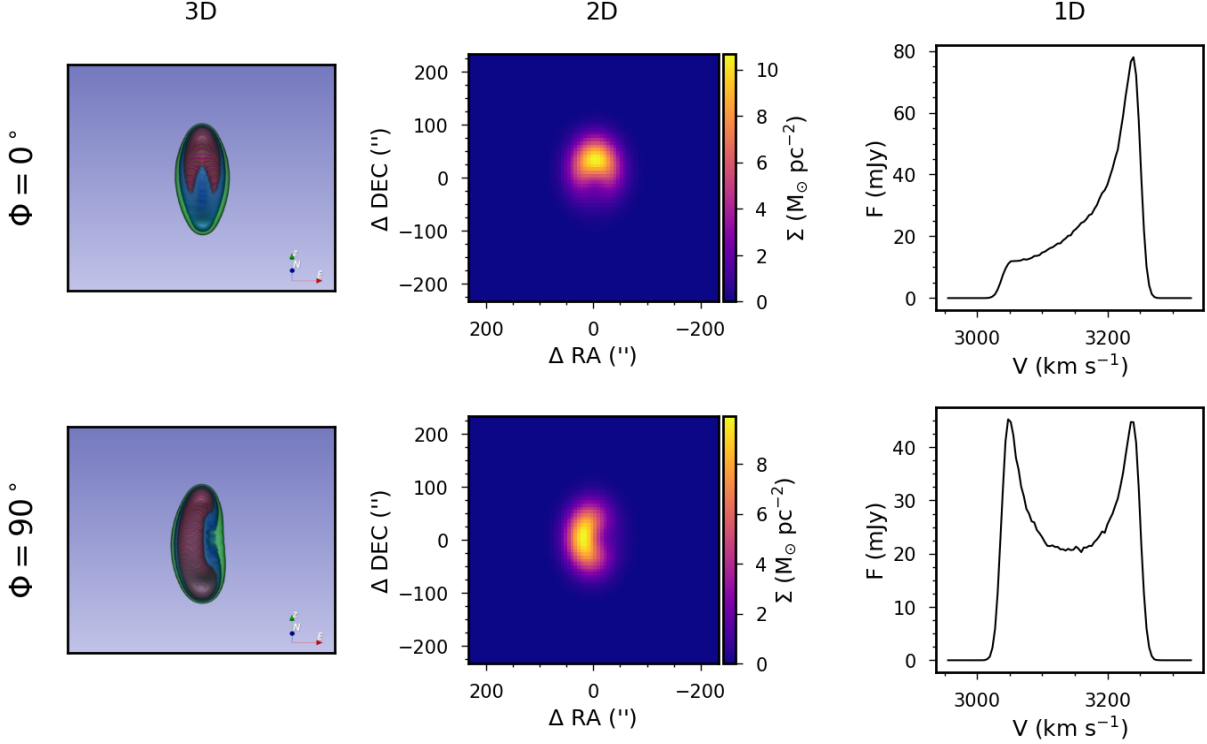}
    \caption{An example of two strongly asymmetric mock cubes generated by our modified version of \textsc{MCGSuite} \textcolor{black}{with different viewing angles, $\Phi$, to the asymmetric feature}. \textcolor{black}{Both rows show a model galaxy with $M_{HI}=10^{9.5}$ M$_{\odot}$, $D_{\hi}=5$ beams across the major axis, $i=45^{\circ}$, $\phi=0^{\circ}$, and $A_{1}=0.8$.  The left-hand panels show 3D projections of the cubelet taken from \textsc{SlicerAstro} that have been oriented to roughly match the moment 0 maps \textcolor{black}{in the middle panels}.  The axes shown in \textcolor{black}{the left} panels as $E,N,Z$ correspond to RA, DEC, and $V_{\rm{los}}$. The colours show surfaces of constant flux in the \textcolor{black}{3D} cubelets.}
    }
    \label{Fig:ExampleCube}
\end{figure*}

We have modified \textsc{MCGSuite} to include an arbitrary number of Fourier moments in the gaseous surface density.  By using the first moment, we are able to generate asymmetric \hi\ cubes for testing purposes.  The strength of this moment is characterized by the $A_{1}$ Fourier \textcolor{black}{coefficient} and can be oriented at an arbitrary phase angle, $\Phi$, measured relative to the major axis of the galaxy.  Explicitly, the galaxy plane surface density is set to
\begin{equation}
    \Sigma(R,\theta)=\Sigma(R)\left(1+A_{1}\cos(\theta+\Phi)\right)~,
\end{equation}
where $\Sigma(R)$ is the \textcolor{black}{axially symmetric} surface density at the cylindrical radius $R$, and $\theta$ is the cylindrical angle in the galaxy plane measured from the approaching side of the major axis.  \textcolor{black}{As illustrated in Figure \ref{Fig:ExampleCube} which shows a pair of example cubelets, $\Phi=0^{\circ}$ corresponds to an asymmetry about the minor axis, while $\Phi=90^{\circ}$ is an asymmetry about the major axis}. \textcolor{black}{For this section, all mock cubelets are noiseless and have $\phi=0^{\circ}$, as the \textcolor{black}{disk} position angle does not affect the calculated asymmetry.}

Figure \ref{Fig:ExampleCube} shows two example cubelets generated by our modified version of \textsc{MCGSuite}.  Both cubes are built with the same underlying model \textcolor{black}{($M_{\hi}=10^{9.5}$ M$_{\odot}$, $D_{\hi}=5$ beams, $i=45^{\circ}$, and $\phi=0^{\circ}$)} and Fourier moment ($A_{1}=0.8$).  The only difference between the cubes is \textcolor{black}{$\Phi$}, the orientation of the asymmetry with respect to the observer's line-of-sight.  The 3D visualizations shown in the left-hand column are generated using the \textsc{SlicerAstro}\footnote{\href{https://github.com/Punzo/SlicerAstro}{https://github.com/Punzo/SlicerAstro}} software package \citep{Punzo2016,Punzo2017}.  These two examples are clearly unrealistic in terms of their level of asymmetry.  However, they show how the orientation of an asymmetric feature in the disk surface density affects the observed cubelet, moment map, and velocity profile.  In particular, the $\Phi=90^{\circ}$ model has a completely symmetric velocity profile, which is consistent with findings of \citet{Deg2020}.

\textcolor{black}{Throughout this paper} we use a newly developed code called \textsc{3DACS}\footnote{\href{https://github.com/NateDeg/3DACS}{https://github.com/NateDeg/3DACS}} (3D Asymmetries for data CubeS) to calculate asymmetries.  A brief description of this publicly available code is given in the appendix.  For this section we only utilize the squared difference asymmetry.  To keep the notation clean, we simply use $A_{3D}$, $A_{2D}$, and $A_{1D}$ \textcolor{black}{in what follows} to represent the 3D, 2D, and 1D `squared difference' asymmetry respectively\textcolor{black}{, computed using Eq.~\ref{Eq:Asq}}.  \textcolor{black}{Furthermore, all analysis here uses the dynamical center of the cube as the rotation point.  Calculating the asymmetry at the dynamical center allows for the strongest linking \textcolor{black}{between} the measured asymmetry to the structure and disruption of a galaxy.}

\subsection{\textcolor{black}{Asymmetry Viewing Angle}}

As seen in Fig. \ref{Fig:ExampleCube}, the orientation \textcolor{black}{$\Phi$} of an asymmetric feature with respect to the line-of-sight can strongly affect the observed morphology.  Unlike other potential observational biases, like inclination and resolution, the viewing angle of an asymmetry is effectively unknown for any galaxy.  While one can select a sample of galaxies based on inclination, resolution, $S/N$, and other effects to avoid potential biases, this is impossible for the viewing angle.  Any survey will contain galaxies with \textcolor{black}{a range} of viewing angles for the intrinsically asymmetric features.

\begin{figure}
    \centering
    \includegraphics[width=\columnwidth]{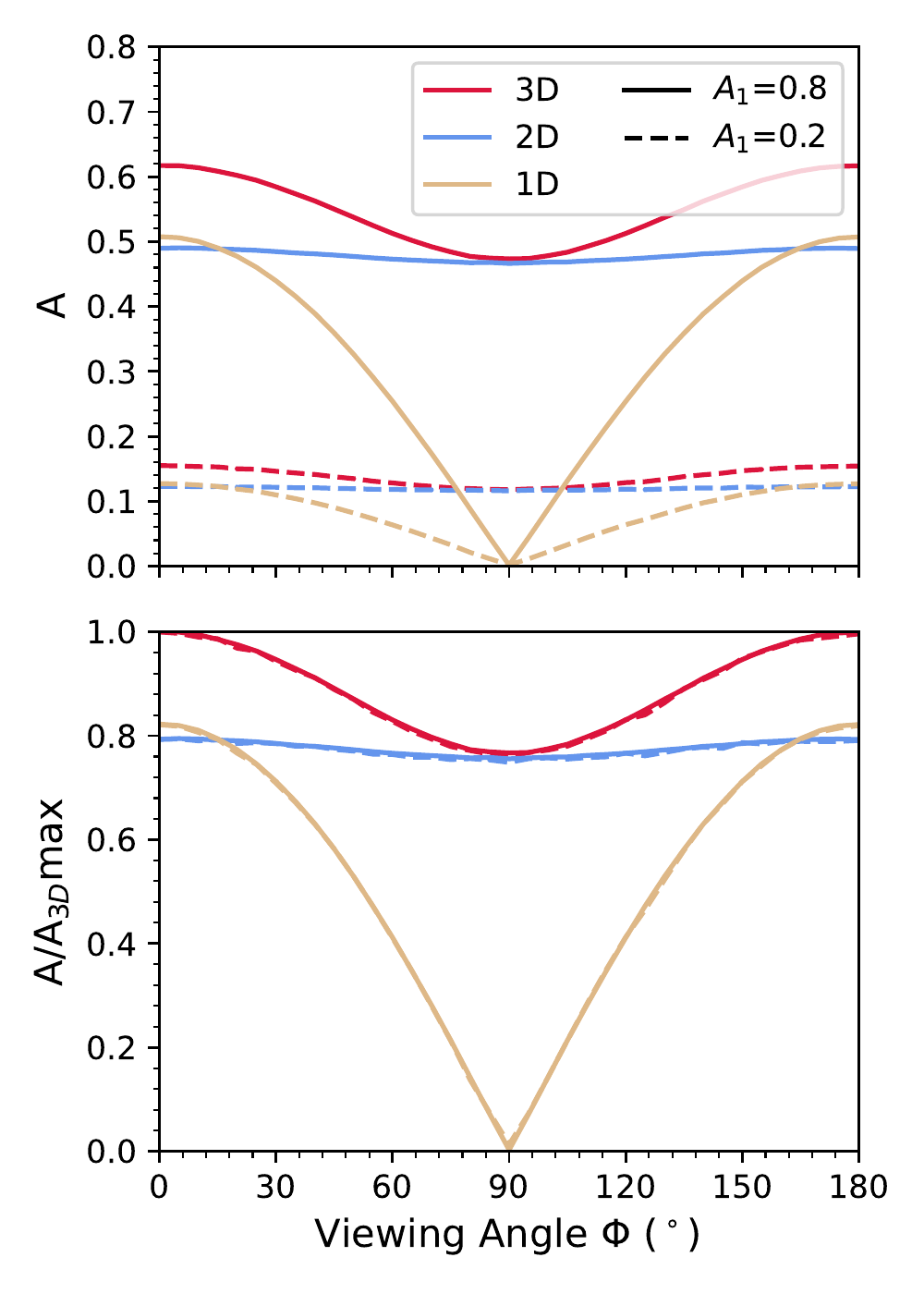}
    \caption{The dependence of the asymmetry measurement on the viewing angle, $\Phi$, of the asymmetric feature.  \textcolor{black}{When $\Phi=0^{\circ}$ or $180^{\circ}$ the asymmetry appears across the minor axis (see the top row of Figure \ref{Fig:ExampleCube}, while when $\Phi=90^{\circ}$ the asymmetry appears across the major axis.}The top panel shows the measured asymmetry while the bottom panel shows the asymmetry scaled to the maximum 3D asymmetry \textcolor{black}{for that particular suite}, $A_{3Dmax}$ for the strongly asymmetric ($A_1 = 0.8$, solid lines) and weakly asymmetric ($A_1 = 0.2$, dashed lines) suites.  Note that in the bottom panel the dashed lines and solid lines are superimposed. \textcolor{black}{All models in this plot have $M_{\hi}=10^{9.5}$ M$_{\odot}$, $D_{\hi}=8$ beams, and $i=50^{\circ}$.}}
    \label{Fig:ViewingAngle}
\end{figure}

Figure \ref{Fig:ViewingAngle} explores the dependence on viewing angle in greater detail.  It shows two suites of noiseless cubes, one with $A_{1}=0.8$ (solid lines) and one with $A_{1}=0.2$ (dashed lines), where $\Phi$ is varied between $0^{\circ}$ and $180^{\circ}$. The base model for each suite has a size, $D_{\hi}$, of 8 beams across and $M_{HI} = 10^{9.5}$ M$_{\odot}$.  The mock galaxies are all observed at an inclination of $i=50^{\circ}$.  
The upper panel of Fig. \ref{Fig:ViewingAngle} shows the calculated asymmetry for each cube, while the bottom panel shows the asymmetry scaled to the maximum asymmetry $A_{3Dmax}$ for that particular suite.  

It is clear that 1D asymmetry is particularly susceptible to the viewing angle, \textcolor{black}{with $A_{1D} = 0$ for $\Phi = 90^\circ$}, which is consistent with the results of \citet{Deg2020}.  The 2D asymmetry displays a greater resilience, remaining nearly constant regardless of the viewing angle.  This can be understood when comparing \textcolor{black}{panels in} the middle column of Fig. \ref{Fig:ExampleCube}: the change in the viewing angle at this inclination does move flux around in the moment map, but it does not affect how asymmetric the image appears.  By contrast, the 3D asymmetry's susceptibility to viewing angle effects lies between these two extremes. The reason for a variation in the 3D asymmetry signal is due to its additional dependence on the velocity structure.  

The bottom panel \textcolor{black}{of Fig.~\ref{Fig:ExampleCube}}, which shows the asymmetry effects of viewing angle scaled by the peak 3D asymmetry, illustrates that once scaled, the asymmetry effects of viewing angle are the same across different input asymmetry amplitudes. The invariance of the scaled asymmetry across different input amplitudes is maintained for inclination and resolution (when dealing with noiseless data).  As such, we have chosen to use unrealistically asymmetric galaxies \textcolor{black}{($A_1=0.8$)} in the noiseless data tests \textcolor{black}{that follow} in order to emphasize the observational effects on the asymmetry measurements.

\subsection{Inclination}

\begin{figure}
    \centering
    \includegraphics[width=\columnwidth]{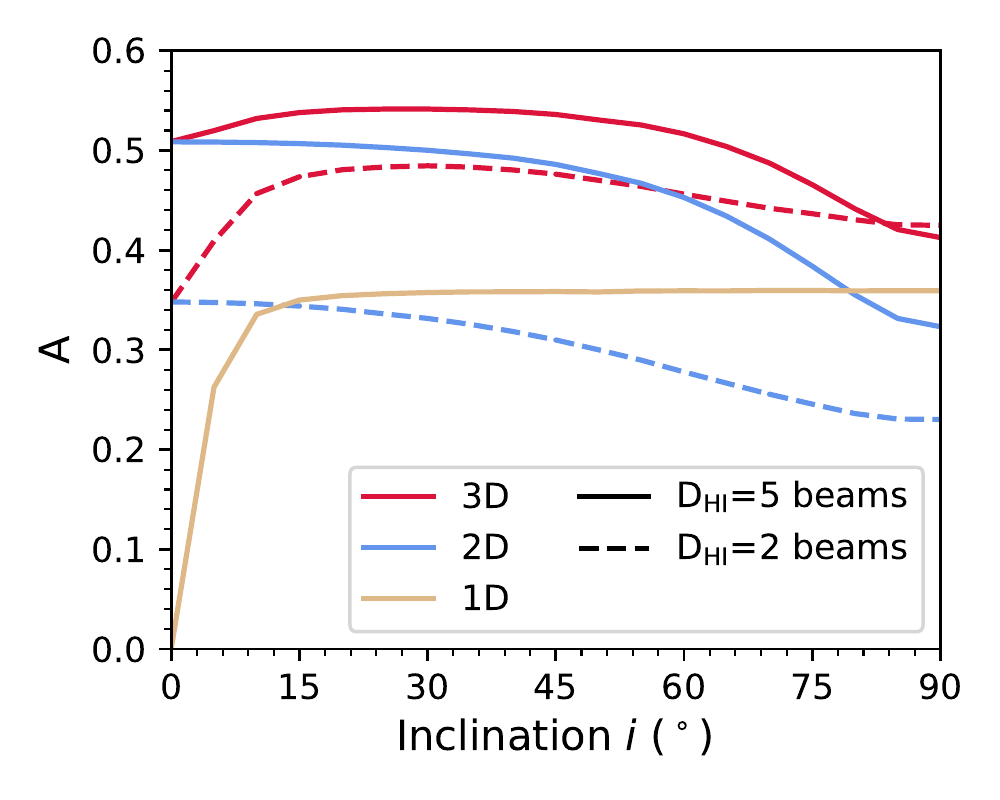}
    \caption{The dependence of the asymmetry on the \textcolor{black}{disk} inclination \textcolor{black}{for moderately-resolved models ($D_{HI} = 5$ beams, solid lines) and marginally-resolved models ($D_{HI} = 2$, dashed lines)}. 
    \textcolor{black}{Only one line is seen for the 1D asymmetry, as the velocity profile does not depend on the spatial resolution.} \textcolor{black}{All models in this plot have $M_{\hi}=10^{9.5}$ M$_{\odot}$, $\Phi=45^{\circ}$, and $A_{1}=0.8$}.}
    \label{Fig:Inclination}
\end{figure}

Inclination must also affect the measured asymmetry of an object.  Figure \ref{Fig:Inclination} shows how the asymmetry varies as a function of \textcolor{black}{disk} inclination.  In this case, the two models shown both have $A_{1}=0.8$, $M_{\hi}=10^{9.5}$ M$_{\odot}$, and \textcolor{black}{an asymmetry} viewing angle of $\Phi=45^{\circ}$, but one model is \textcolor{black}{moderately resolved with $D_{HI} = 5$} beams across while the other \textcolor{black}{is} only \textcolor{black}{marginally resolved with $D_{HI} = 2$} beams across.  

This figure shows that a face-on galaxy will have greatly diminished 1D asymmetry whereas an edge-on galaxy will have a reduced 2D asymmetry. Extreme inclinations also \textcolor{black}{lower the} 3D asymmetry, but it always \textcolor{black}{includes the signal from the} 2D or 1D asymmetry, making it more resilient to inclination effects than either statistic \textcolor{black}{alone}. Fig.~\ref{Fig:Inclination} also illustrates how at lower spatial resolutions, 3D asymmetry \textcolor{black}{more closely resembles the} 1D asymmetry.  Additionally, at low inclinations where the 1D asymmetry goes to zero, the 3D asymmetry reduces to the 2D asymmetry.

Another important takeaway from Fig. \ref{Fig:Inclination} is that the \textcolor{black}{relative shapes of the inclination trends are not constant with resolution.} This is different than the response of the asymmetry to different Fourier strengths seen in Fig. \ref{Fig:ViewingAngle}.  Therefore, understanding the effect of the resolution on the measured asymmetry is critically important.

\subsection{Resolution}\label{subsec:Resolution}

\begin{figure}
    \centering
    \includegraphics[width=\columnwidth]{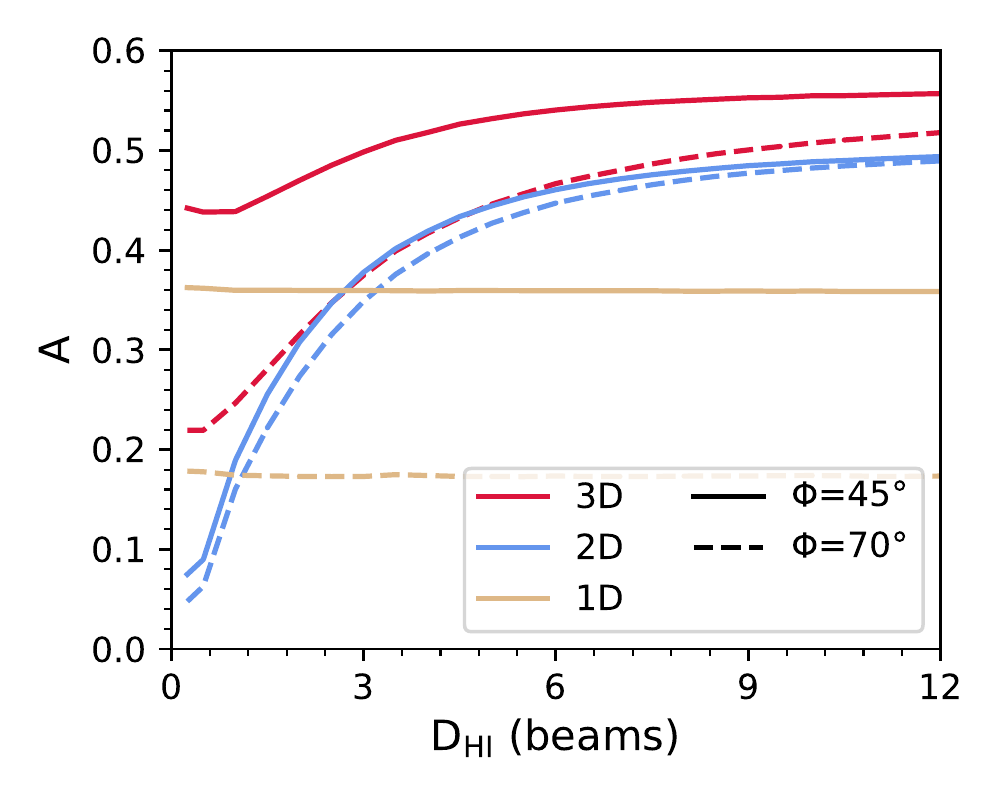}
    \caption{The dependence of the asymmetry on the spatial resolution at two different viewing angles \textcolor{black}{(solid and dashed lines)}. \textcolor{black}{All models in this plot have $M_{\hi}=10^{9.5}$ M$_{\odot}$, $i=50^{\circ}$, and $A_{1}=0.8$}.}
    \label{Fig:Resolution}
\end{figure}

To understand the effect of \textcolor{black}{spatial} resolution on the measured asymmetry, an additional two suites of model cubes were generated with varying 
$D_{\hi}$,
\textcolor{black}{with} $M_{HI}=10^{9.5}$ M$_{\odot}$ and $i=50^{\circ}$. One set of models has $\Phi=45^{\circ}$, while the other has $\Phi=70^{\circ}$. Figure \ref{Fig:Resolution} shows the dependence of asymmetry \textcolor{black}{in these two suites} as a function of the spatial resolution. The 1D asymmetry, which is purely spectral, does not depend on the spatial resolution, but it does depend on the viewing angle. By contrast the 2D asymmetry plunges at lower resolutions, and should converge to zero when the object is effectively unresolved.  The 3D asymmetry, though affected by resolution, remains more resilient to the spatial resolution than 2D asymmetry.  When the model becomes unresolved, the 3D asymmetry should converge to the 1D asymmetry.

Taking the effects of viewing angle, inclination, and particularly resolution together, it is clear that in the marginally-to-moderately resolved regime, the 3D asymmetry is \textcolor{black}{more representative of the intrinsic asymmetry than} either the 1D or 2D asymmetries.  The 3D asymmetry is less susceptible to viewing angle variations than the 1D asymmetry, is more resilient to inclination effects than either the 1D or 2D measures, and can be used at lower resolutions than the 2D asymmetry.  

A key point to note is that, while the observed geometry/resolution affects all asymmetry measurements, an important use of any asymmetry statistic is separating surveys of galaxies into undisturbed and disturbed populations.  While none of the statistics are constant with geometry/resolution, the 3D asymmetry \textcolor{black}{shows a variation maximum of 0.25 in Fig. \ref{Fig:Resolution}, and an average variation across Figs. \ref{Fig:ViewingAngle}-\ref{Fig:Resolution} of $\approx 0.05$, which are both \textcolor{black}{lower} than the variations seen for the 2D and 1D statistics.  This lower variation} implies that it will be less likely to classify a disturbed galaxy as undisturbed due to the orientation of the galaxy with respect to the observer.  While this section has only used the `squared difference' asymmetry statistic, when this analysis is repeated with the traditional `absolute difference' asymmetry the results are the same.

\section{Uncertainties and Noise}\label{Sec:Noise}

\subsection{Uncertainties}\label{subsec:Uncertainty}

\textcolor{black}{In addition to biasing the asymmetry measure itself (see Section~\ref{subSec:AsymNoise}), noise also adds random uncertainty to the asymmetry that must be calculated.  This uncertainty arises in three distinct ways: through the formal uncertainties from the noise, by causing variations in the mask, and through uncertainties in the precise rotation point.  By definition, the formal \textcolor{black}{uncertainty $\sigma_{A_{sq}}$ is given by:}}
\begin{equation}\label{Eq:Sigma}
    \sigma_{A_{sq}}^{2}=0.5A_{sq}^{2}\left(\frac{\sigma_{P_{c}}^{2}}{P_{c}^{2}}+ \frac{\sigma_{Q_{c}}^{2}}{Q_{c}^{2}}\right)~,
\end{equation}
\textcolor{black}{where $P_{c}=P_{sq,m}-B_{sq}$ and $Q_{c}=Q_{sq,m}-B_{sq}$.  Calculating this uncertainty is not trivial due to the signal-noise cross terms in $P_{sq,m}$ and $Q_{sq,m}$ seen in Eq. \ref{Eq:PExpand}.  While the expectation value of those terms is zero, they nonetheless introduce uncertainty.}

\textcolor{black}{It is possible \textcolor{black}{to write an expression for $\sigma_{A_{sq}}$ starting from Eq.~\ref{Eq:Sigma} and applying a number of approximations that simplify it to some degree}.  However, in practice, this uncertainty is small relative to the systematic uncertainties.  \textcolor{black}{For example, in} our tests we found that the formal uncertainty \textcolor{black}{computed from Eq.~\ref{Eq:Sigma}} is rarely larger than $\sigma_{A_{sq}}=0.02$.  By comparison, the uncertainty associated with the unknown viewing angle shown in Figure \ref{Fig:ViewingAngle} is on the order of $0.1$. \textcolor{black}{We elaborate on the magnitude of different sources of uncertainty on $A_{sq}$ below.}}

\begin{figure}
    \centering
    \includegraphics[width=\columnwidth]{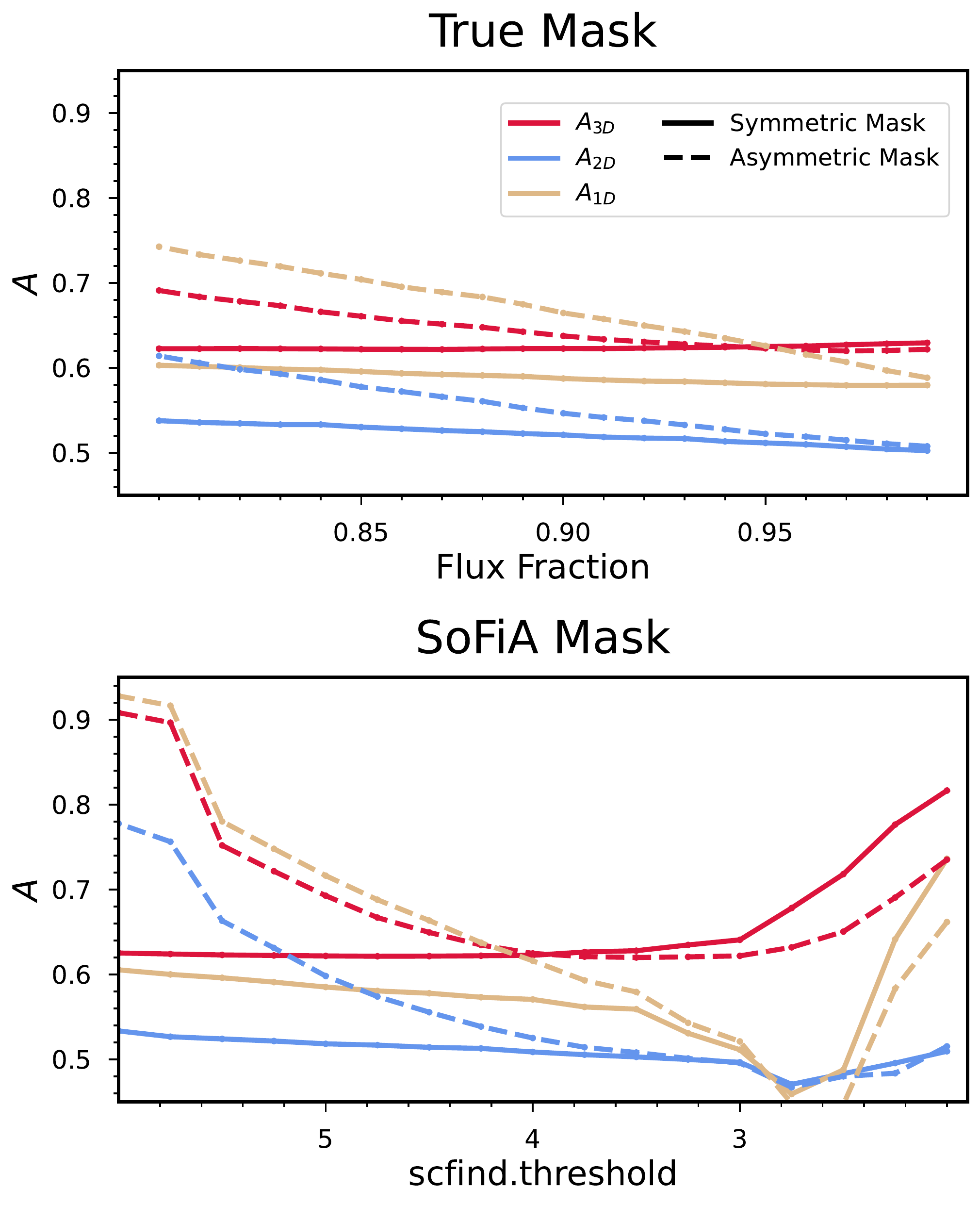}
    \caption{The noise corrected asymmetry measured for a cube with an input noise of \textcolor{black}{1.6 mJy per 30\arcsec\ beam} and $M_{\rm{HI}}=10^{9.5}~\rm{M}_\odot$, $D_{\rm{HI}}=5$ beams, $i=45^{\circ}$, $A_{1}=0.8$.  The top panel shows the asymmetry using masks constructed from the underlying noiseless cube where the x-axis is the \textcolor{black}{fraction of the} total noiseless flux included in the mask.  The bottom panel uses masks generated by \textsc{SoFiA-2} with different values of the \textsc{SoFiA-2} parameter \textit{scfind.threshold} \textcolor{black}{either as is (Asymmetric mask, dashed lines) or symmetrized as described in the text (Symmetric mask, solid lines)}.  In both panels the total flux included in the mask increases towards the right.}
    \label{Fig:Mask}
\end{figure}

\textcolor{black}{Beyond the formal uncertainty, noise can generate variations in the masks/segmentation maps that are used in the asymmetry calculation.  The construction of such masks is not trivial and variations in the mask due to noise may affect the asymmetry in non-obvious ways.  In order to explore the effect of mask variations on the asymmetry measurement only, we added Gaussian noise with \textcolor{black}{$\sigma=1.6$ mJy per 30\arcsec\ beam (as expected for the WALLABY \hi\ survey; \citealt{Koribalski2020})} to the \textcolor{black}{cubelet shown} in the upper row of Figure \ref{Fig:ExampleCube} ($M_{\rm{HI}}=10^{9.5}~\rm{M}_\odot$, $D_{\rm{HI}}=5$ beams, $i=45^{\circ}$, $\phi=0^{\circ}$, $A_{1}=0.8$).  Figure \ref{Fig:Mask} shows the measured asymmetry in 1D, 2D, and 3D for masks constructed using a fraction of the total flux of the noiseless cube (top panel), and masks constructed using \textsc{SoFiA-2} with different source finding thresholds, \textcolor{black}{where the total cube flux included by the mask increases towards low values of {\it scfind.threshold}.} (bottom panel).  In both panels the dashed lines show the effect of using these unmodified and asymmetric masks on the measured asymmetry. The variations in the dashed lines show that, when using asymmetric masks, the precise size and shape of the mask can \textcolor{black}{change the uncertainty by tens of percent for relatively modest changes in source finding parameters.}}

\textcolor{black}{An approach to mitigate this effect is to \textit{symmeterize} the \textcolor{black}{mask within which the asymmetry is calculated}.  This is done by adjusting the input mask such that pairs of cells/pixels/channels across the rotation point are either both \textcolor{black}{included in or excluded from} the mask.  The solid lines of Figure \ref{Fig:Mask} show that the variations in the measured asymmetry are much smaller when using these symmetric masks.  There are some variations \textcolor{black}{when the sourcefinding threshold digs deeper into the noise ({\it scfind.threshold} $< 3$).}
\textcolor{black}{The resulting} large masks are including a great deal of additional flux \textcolor{black}{from noise peaks, and are} unlikely to be used for real observations.}  

\textcolor{black}{Given that the asymmetry variations from symmetric masks \textcolor{black}{with ({\it scfind.threshold} $> 3$)} are $\lesssim 0.02$ \textcolor{black}{(solid lines in Fig.~\ref{Fig:Mask})}, we utilize symmetric masks throughout this work.  However, it should be noted that it is not always possible to construct a symmetric mask, and in such cases it will be necessary to consider how to estimate the asymmetry uncertainty due to potential mask variations.  For instance, when minimizing the asymmetry for a 2D image or 1D profile calculated from a 3D datacube, symmetrizing the mask is impossible because the image and profile are generated using a 3D mask.}

\textcolor{black}{Yet another way that noise may affect the asymmetry is \textcolor{black}{by introducing} uncertainties in the rotation point.  Noise can cause uncertainties in the measurement of the dynamical center (or other interesting pivot points) which should propagate to an uncertainty in the measured asymmetry.  The simplest way to propagate this uncertainty is to simply calculate the asymmetry \textcolor{black}{within the range of allowable points given the uncertainty} and use the extrema to determine the asymmetry uncertainty.  In 3D \textcolor{black}{for uncorrelated uncertainties}, this involves calculating the asymmetry at an additional 26 points; these are the $3\times 3\times 3-1$ points without the center point defined by $(x\pm \delta x,y \pm \delta y,v\pm \delta v)$.  To give an idea of the scale of these variations we assumed an uncertainty of $\pm 0.25$ beams and $\pm 1$ channels \textcolor{black}{(typical for \hi\ datacubes from widefield surveys, e.g.\ \citealt{Deg2022})} for the center of the mock cube used in the mask tests \textcolor{black}{shown in Fig.~\ref{Fig:Mask}}.  Setting the uncertainty as half the range of the minimum and maximum uncertainties gives $\sigma_{A,\rm{center}}=0.1$.  While this example is informative, the precise size of this variation will strongly depend on how precisely the rotation point is known.  For instance, if the optical center of brightness is used, the uncertainty on the rotation point will likely be much lower.}

\textcolor{black}{Table \ref{tab:ExUncertainties} lists the three different sources of uncertainty for the mock cube in Fig~\ref{Fig:Mask} as well as the background contribution to the asymmetry.  It also includes the actual measured values for $P_{sq,m}$ and $Q_{sq,m}$ as $B_{sq,3D}$ appears in both the numerator and denominator of the asymmetry calculation.  In this example, the dominant uncertainty is the  uncertainty \textcolor{black}{on the rotation point, which we have assumed to be} $\pm 0.25$ beams and $\pm 1$ channel.  It is worth noting that the formal uncertainty (Eq.~\ref{Eq:Sigma}) is considerably smaller than the variations due to the viewing angle seen in Figure \ref{Fig:ViewingAngle}.  As discussed in Sec. \ref{Sec:Geometry}, the viewing angle is an uncontrollable and unknowable parameter from an observational point of view.  Thus, when observing a population, the systematic uncertainty in the measurements from such observational biases will likely dominate over other sources of uncertainty. } 


\begin{table}
\begin{tabular}{|l|c|}
    \hline 
     Measurement & Value \\
     \hline 
     $A_{3D}$ &  0.63\\
     $P_{sq,m}$ & 0.54\\
     $Q_{sq,m}$ & 1.19\\
     $B_{sq,3D}$ & 0.11\\
     $\sigma_{A_{sq}}$ & $\sim0.02$\\
     Symmetric Mask Uncertainty & 0.02\\
     Rotation Point Uncertainty & 0.08\\
     \hline 
\end{tabular}
\label{tab:ExUncertainties}
\caption{\textcolor{black}{Sources of uncertainty for \textcolor{black}{the noisy cube used to generate Fig.~\ref{Fig:Mask}}: noise of 1.6 mJy per 30\arcsec\ beam, and $M_{\rm{HI}}=10^{9.5}~\rm{M}_\odot$, $D_{\rm{HI}}=5$ beams, $i=45^{\circ}$, $A_{1}=0.8$.  An uncertainty in the rotation point of $\pm 0.25$ beams and $\pm 1$ channel was assumed.}}
\end{table}


\subsection{Squared Difference vs Absolute Asymmetry}

The effect of noise on asymmetry measurements has been investigated in a large number of works.  As noted in \citet{Conselice2000} and \citet{Conselice2003}, noise will always increase the measured value of the asymmetry.  To account for this when using the `absolute difference' asymmetry, a background measurement is made and subtracted from the measured value as shown in Eq. \ref{Eq:AbsAsym_Background}.  However, as noted in \citet{Giese2016} and \citet{Thorp2021}, this subtraction is an approximation and will overcorrect the asymmetry down to zero in the low $S/N$ regime.  In Sec. \ref{subSec:AsymNoise}, we introduced the `squared difference' asymmetry that should have a more robust background subtraction. 

\begin{figure*}
    \centering
    \includegraphics[width=0.9\textwidth]{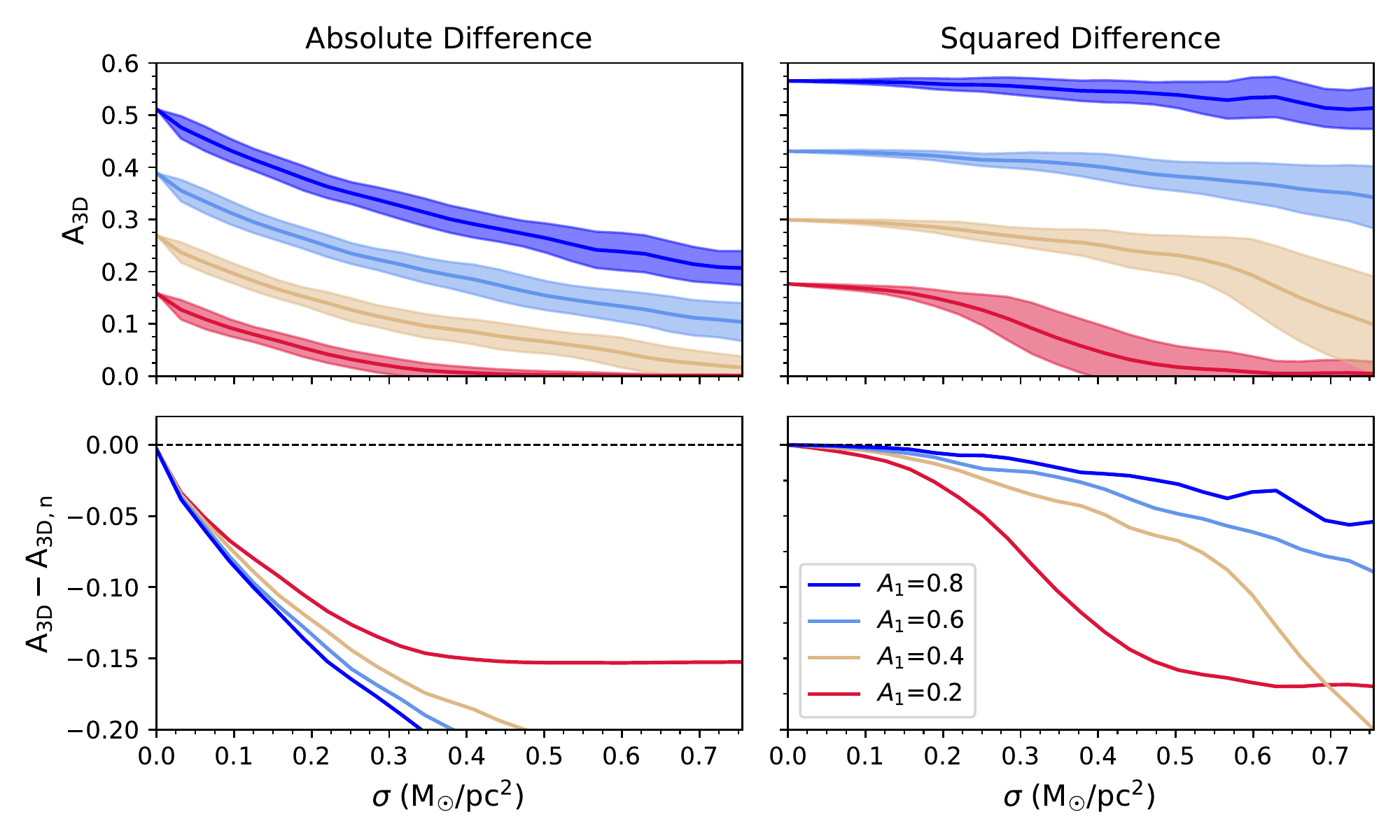}
    \caption{\textcolor{black}{Behaviour of the asymmetry statistic as a function of cubelet noise for different input Fourier moments \textcolor{black}{in the range $0.2 \leq A_1 \leq 0.8$, when asymmetries are calculated using `absolute difference'  and `squared difference' methods.}} 
    In the top row, the solid lines show the average value calculated using a Gaussian kernel, and the shaded regions show the standard deviation of points about the mean.  Negative \textcolor{black}{absolute asymmetry} values obtained from over-subtraction are treated as 0. The bottom row shows the difference between the measured asymmetry, $A_{3D}$, and asymmetry from the noiseless cubes, $A_{3D,n}$. \textcolor{black}{The dashed black line at 0 highlights what is expected when the background corrected asymmetry matches the noiseless asymmetry.} \textcolor{black}{All models used to generate these curves have $M_{\hi}=10^{9.5}$ M$_{\odot}$, $D_{\hi}=8$ beams, $i=50^{\circ}$, and $\Phi=45^{\circ}$}.}
    \label{Fig:SN}
\end{figure*}

In order to compare the effect of \textcolor{black}{the} noise \textcolor{black}{contribution to the asymmetry} \textcolor{black}{when `absolute differences' and when `squared differences' are used}, we built a suite of 1000 cubelets \textcolor{black}{with} increasing levels of noise, $\sigma$, ranging from \textcolor{black}{0.01~mJy to 5 mJy per 30\arcsec\ beam}.  Each cube has $M_{\hi}=10^{9.5}$ M$_{\odot}$, $i=50^{\circ}$, and $\Phi=45^{\circ}$.  Half the cubes have $A_{1}=0.8$ while the other \textcolor{black}{half} have $A_{1}=0.2$. Once generated, \textsc{SoFiA-2} was used to make a mask for each cube, \textcolor{black}{within which the asymmetry was then calculated about the centre of the axisymmetric component of the mock \hi\ disk}.  
The upper panels of Figure \ref{Fig:SN} show the background-corrected asymmetry for each statistic \textcolor{black}{as a function of the input noise \textcolor{black}{expressed in $\rm{M}_{\odot}\,\rm{pc}^{-2}$ in a $30\arcsec$ beam over a spectral range of 16 $\kms$ ($\sim$ 4 WALLABY spectral channels)}}, while the lower panels show the difference between the measured asymmetry and the asymmetry calculated from a noiseless cube, $A_{3D,n}$. 
For the upper panels, we show an average asymmetry and a width of one standard deviation \textcolor{black}{at each noise value calculated using a Gaussian kernel with a width that is inversely proportional to the density of points.}


Figure \ref{Fig:SN} shows the general bias of the `absolute difference' asymmetry quite clearly.  \textcolor{black}{There is a constant decrease in the asymmetry, with a slope that is roughly independent of the initial Fourier moment strength (as seen in the lower left panel).  There is a turnover as the noise corrected `absolute difference' asymmetry reaches zero, as seen in the red lines.  By $\sigma=0.25\,\rm{M}_{\odot}\,\rm{pc}^{-2}$, the measured asymmetry has decreased by 0.15.  By contrast the `squared difference' asymmetry remains at a constant value until $\sigma\approx 0.2-0.3\,\rm{M}_{\odot}\,\rm{pc}^{-2}$.  Both the start of and the rate of the decrease depends somewhat on the initial strength of the Fourier moment.
Interestingly, the spread of measured asymmetry values is roughly constant for the `absolute difference' statistic, while it increases with noise for the `squared difference' statistic.  In the region of the plot where the `squared difference' asymmetry has low bias, the standard deviation is below 0.01. }

\begin{figure*}
    \centering
    \includegraphics[width=0.9\textwidth]{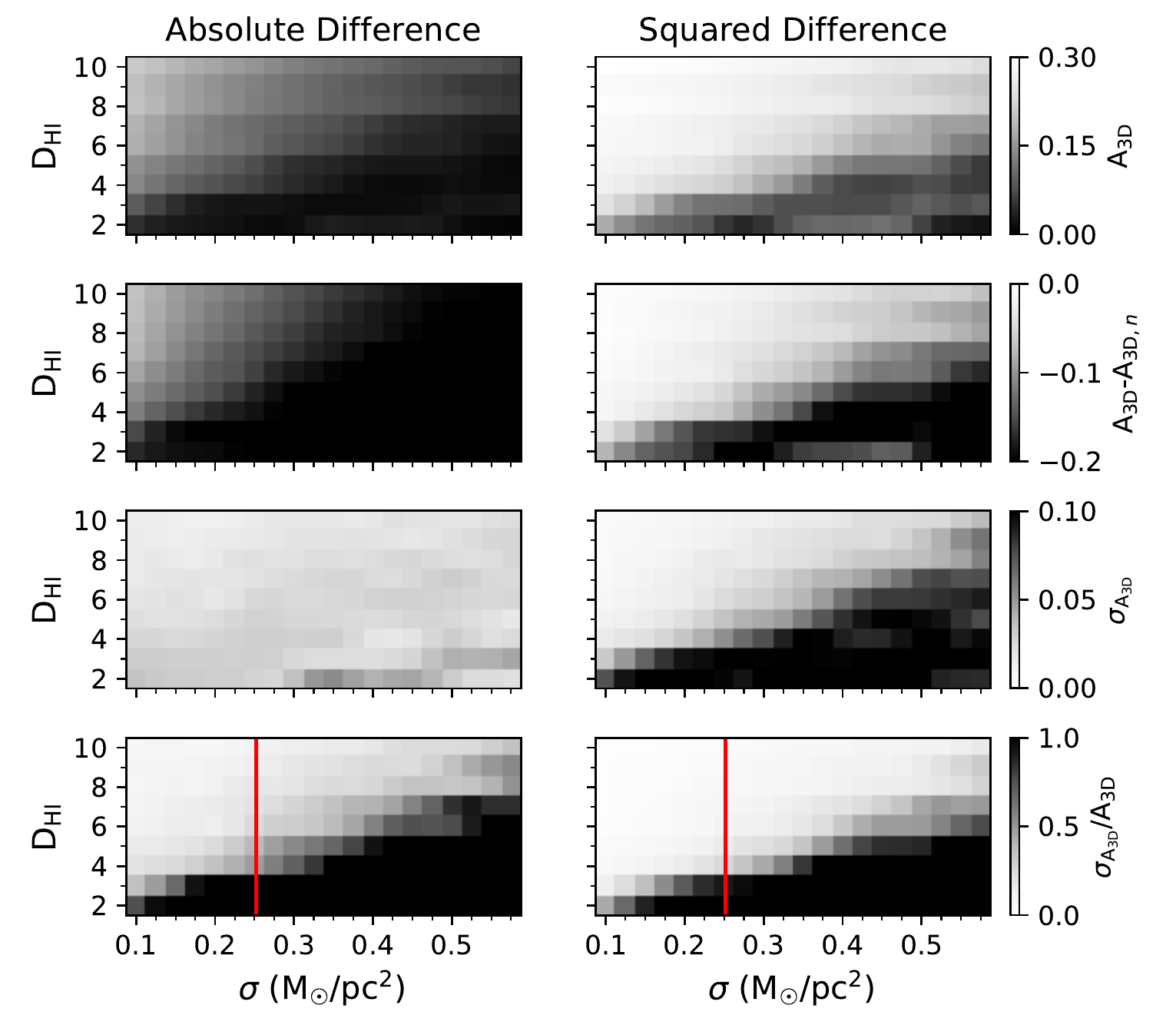}
    \caption{\ Intensity maps portraying the effects of resolution and noise on the absolute difference asymmetry and squared difference asymmetry measurements for a galaxy.  The units of $D_{\hi}$ are beams.  The upper panels show the average background corrected 3D asymmetry \textcolor{black}{while the second row shows the difference between this background corrected asymmetry and the asymmetry of an equivalent noiseless cubelet.} The third row shows the spread in asymmetries in each bin, which is calculated as the standard deviation of all points in each cell in the $\sigma_-D_{\hi}$ space. \textcolor{black}{The bottom row panels show the relative spread in the asymmetry measurement.}  \textcolor{black}{The vertical line shows the noise for a WALLABY-like population (see Sec. \ref{Sec:WallabyExample} for a discussion of this population.)} \textcolor{black}{All models used to generate this map have $M_{\hi}=10^{9.5}$ M$_{\odot}$, $i=50^{\circ}$, $\Phi=45^{\circ}$, and $A_{1}=0.4$}.  }
    \label{Fig:ResSNMap}
\end{figure*}

The effect of the noise on the asymmetry measurement is also related to the \textcolor{black}{spatial} resolution.  To demonstrate this, Figure \ref{Fig:ResSNMap} shows the background corrected asymmetry (top row), \textcolor{black}{the difference between the corrected and noiseless asymmetries (second row)}, the measured spread (third row), and relative uncertainty (bottom row) for a suite of cubes with different resolutions and input levels of noise.  Each cube has $M_{\hi}=10^{9.5}$ M$_{\odot}$, $i=50^{\circ}$, $\Phi=45^{\circ}$, and an input Fourier strength of $A_{1}=0.4$.

\textcolor{black}{In both the `absolute difference' and `squared difference' asymmetry, there is a clear dependence on the spatial resolution.  Better-resolved objects are both less biased by the noise than more poorly-resolved objects, and have a lower uncertainty in their measured asymmetry. Objects with $D_{\hi} \geq 8$ beams have only a small bias even at large levels of noise. \textcolor{black}{These results are qualitatively similar to the resolution tests presented for noiseless cubelets in Fig.~\ref{Fig:Resolution}.}}

\textcolor{black}{As in Fig.~\ref{Fig:SN}, Fig.~\ref{Fig:ResSNMap} shows that the `absolute difference' asymmetry has a relatively constant level of spread, regardless of the noise and object size, while the `squared difference' depends on both the noise and resolution.  This uncertainty is due to the formal uncertainty discussed in Sec. \ref{subsec:Uncertainty} and does not include contributions due to uncertainties in the rotation point.  Below a noise limit of $0.25\,\rm{M}\,\rm{pc}^{-2}$ and for $D_{\hi} > 5$ beams, the `squared difference' asymmetry has a lower spread than the `absolute difference'. }


\textcolor{black}{For many science cases, the key quantity is the relative spread, which is shown in the bottom row of Figure \ref{Fig:ResSNMap}.  Here we see that the relative uncertainty of both methods is similar, although the region where the relative uncertainty is minimized is larger for the `squared difference' method.  The similarity between the two is due to the offsetting behaviours of the bias and uncertainties for each method. Knowing the relative spread can help to plan out where measuring a particular asymmetry statistic is viable at both an individual and at a population level.  }

\textcolor{black}{It is useful to consider the mock cubelet noise in the context of current widefield surveys.  To that end, Fig.~\ref{Fig:ResSNMap} shows the expected noise for WALLABY ($1.6$ mJy per 30\arcsec\ beam, \citealt{Koribalski2020})  in $\rm{M_\odot \, pc}^{-2}$ over a spectral range of $16\,\kms$ (= 4 WALLABY channels) as a vertical red line.  At this noise level, the `absolute difference' asymmetry shows a strong bias at all sizes.  However, the `squared difference' asymmetry is shows little to no bias for all detections with $D_{\hi} \gtrsim 6$ beams.  This suggests that, while the `squared difference' statistic can be used for many of the resolved WALLABY detections, the absolute difference method cannot.}

Altogether, Figs. \ref{Fig:SN} and \ref{Fig:ResSNMap} show that the `squared difference' asymmetry is superior to the `absolute difference' asymmetry \textcolor{black}{in the presence of noise}.  
\textcolor{black}{As such, we recommend that any study of asymmetry \textcolor{black}{adopt} squared differences.}

\section{Potential for WALLABY-like observations}\label{Sec:WallabyExample}

There are a variety of new telescopes undertaking cutting edge widefield \hi\ surveys.  For example, WALLABY on the Australian Square Kilometer Array Pathfinder (ASKAP, \citealt{Hotan2021}) telescope will observe the \hi\ content of galaxies over most of the Southern Sky.  The majority of detections in such surveys are marginally resolved and low $S/N$.  For instance, Fig. 1 of \textcolor{black}{\citet{Deg2022}} shows that most of the \textcolor{black}{detections} in the WALLABY Pilot Data Release 1 (PDR1, \citealt{Westmeier2022}) have $\log(S/N)_{\rm{int}} \le 2$ and $\code{ell\_maj} \le 5$ beams, \textcolor{black}{where \code{ell\_maj} is an estimate of the \textcolor{black}{detection} size based on the source finding (for PDR1, $D_{\hi}\approx 2$ \code{ell\_maj}, \citealt{Deg2022}) and $S/N_{\rm{int}}$ is the integrated $S/N$ in the mask}.

Given the performance of the squared difference 3D asymmetry statistic, it is natural to explore whether it can be applied to WALLABY and other \textcolor{black}{similar} surveys.  While there are strong hints that this is possible from Fig. \ref{Fig:ResSNMap}, those maps are made using a single galaxy model observed at different \textcolor{black}{noise levels}, whereas a real survey will make many different \textcolor{black}{detections} with a common level of noise.  To that end, we generated a population of 500 mock \hi\ cubes with random geometries, sizes, and asymmetry Fourier moments.  Each cube is generated with the nominal WALLABY \textcolor{black}{observing parameters} of $1.6$ mJy/beam, a 30\arcsec\ beam, 6\arcsec\ pixels, and 4 $\kms$ channels \citep{Koribalski2020}.  The mock galaxies have $8 \le \log(M_{\hi}/M_{\odot}) \le 10.5$, $0^{\circ}\le i \le 90^{\circ}$, $0^{\circ} \le \phi \le 360^{\circ}$, $1 \le D_{\hi} \le 40$~beams.  The mass for each galaxy is drawn from a logarithmic distribution, while the other parameters are drawn from linear distributions.  These selections and distributions are meant to roughly approximate the observations of \textcolor{black}{WALLABY PDR1 galaxies} except for the asymmetry levels, but they are not precise matches.  For the asymmetries, the models have $0 \le A_{1} \le 0.6$, and $0^{\circ} \le \Phi \le 180^{\circ}$.  The upper limit on the $A_{1}$ range reflects the fact that no real galaxies should have such a large Fourier $A_{1}$ moment.  The final population, while comprised of galaxies with a different observed size and asymmetry distribution, is still similar enough to WALLABY to draw a few conclusions.

We run \textsc{SoFiA-2} on each cube using the \textcolor{black}{parameters listed for the Hydra Team Release 2 sources} in Table 2 of \textcolor{black}{\citet{Westmeier2022}}.  This generates similar masks to the WALLABY PDR1 observations. At low \textcolor{black}{resolutions} \textsc{SoFiA-2} can fail to find the galaxy, or generate a mask that is not quite appropriate.  Since we know the center of the galaxy, we remove all galaxies where the \textsc{SoFiA-2} center differs from the true center by $\ge15\%$ of the size of the object as estimated by \textsc{SoFiA-2}.  This is a rough filter and a number of galaxies with poorly constructed masks still fall into the sample.  \textcolor{black}{As noted earlier, \code{ell\_maj}$\sim D_{HI}/2$ when using WALLABY-like parameters \citep{Deg2022}.}

\begin{figure}
    \centering
    \includegraphics[width=\columnwidth]{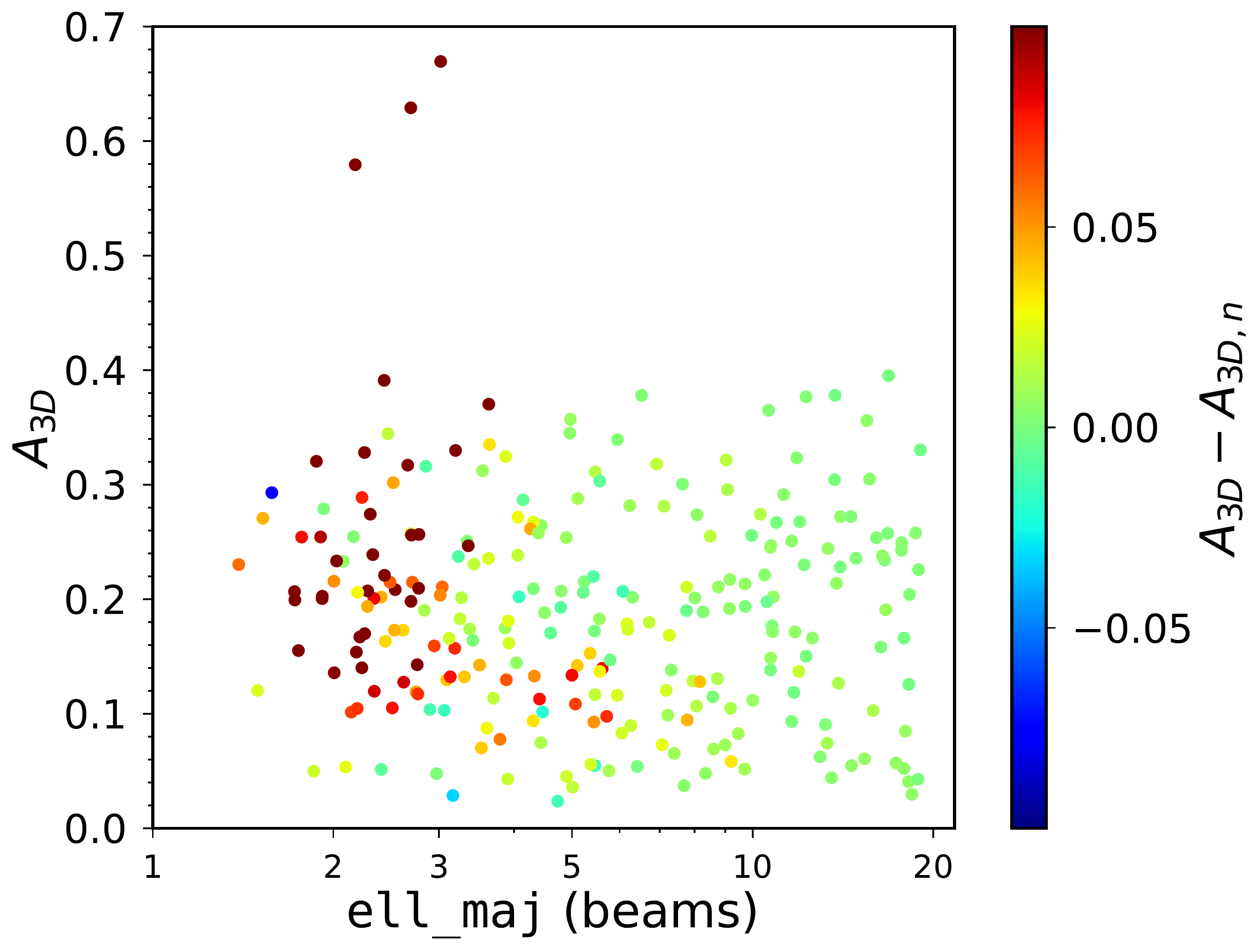}
    \caption{Background-corrected squared difference asymmetry for a randomized population of 500 galaxies \textcolor{black}{detected by a WALLABY-like survey} as a function of their size.  The point colors show the difference \textcolor{black}{between measured asymmetry of the noisy cubes and \textcolor{black}{that of} the noiseless cubes}. }
    \label{Fig:MockWallabyPopAsymDiff}
\end{figure}

Figure \ref{Fig:MockWallabyPopAsymDiff} shows the size, background corrected squared difference asymmetry, and the difference between that measurement and what would be measured for a noiseless cube using the same model and mask.  At lower \textcolor{black}{resolutions} (\code{ell\_maj}$\le 3$ beams), there is still a significant population of galaxies where the mask is poorly constructed.  Above this size, most of the galaxies have well-recovered asymmetries.  \textcolor{black}{This is broadly consistent with the results seen in Figure \ref{Fig:ResSNMap}.  In that figure, the mock galaxies with $D_{\hi}<6$ beams (which is equivalent to \code{ell\_maj}$=3$ beams) show a significant bias, while those above that limit show very little bias.  }  \textcolor{black}{However, there are a few larger objects with lower measured asymmetries where the background subtraction has undercorrected the results by $0.03-0.07$.  The increased asymmetry of these objects is likely due to poorly constructed masks.  }



In WALLABY, and other \textcolor{black}{wide area} \textcolor{black}{untargeted} surveys, greater care will be taken with detecting sources and constructing the masks than we utilized for this toy problem.  As such, it is likely that the measured asymmetry can still be used with a great degree of accuracy for the entire population of marginally resolved detections.  But, if we are to be cautious, Fig. \ref{Fig:MockWallabyPopAsymDiff} \textcolor{black}{suggests} that the asymmetry is accurately measured for all our mock galaxies with \code{ell\_maj}$\ge 3$ beams.  \textcolor{black}{It is worth noting that this is larger than the limit of $D_{\hi}\ge 3-4$ seen in Figs. \ref{Fig:Resolution} and \ref{Fig:ResSNMap}.  This is due to the low $S/N$ of the WALLABY-like population used in Fig. \ref{Fig:MockWallabyPopAsymDiff}.  Nonetheless, even when restricted to \code{ell\_maj}$\ge 3$, the 3D squared difference asymmetry can be applied to the majority of the marginally resolved WALLABY PDR1 detections without worry of noise biasing the results of the analysis, and with little scatter from what one would expect from a noiseless measurement.}

\section{Discussion and Conclusions}\label{Sec:Conclusion}

In this work we have introduced a methodology for calculating the 3D asymmetry of a cubelet containing a single galaxy.  While this method has been designed for \hi\ datacubes, it should be applicable to any spectral line cube, whether from an IFU or using some other line/feature.

The 3D asymmetry is less affected by the viewing angle of the asymmetric feature than the 1D measurement.  It is also superior to both the 1D and 2D measures with respect to the inclination of the galaxy.  The 3D asymmetry can be used at lower spatial resolutions than the 2D measurement.  This result is of particular importance as there are usually an order of magnitude more galaxies that are marginally resolved than are well resolved \textcolor{black}{in widefield, untargeted \hi\ surveys} \citep{Koribalski2020}. \textcolor{black}{The application of asymmetries to large surveys is the key use of this statistic. On an individual basis, the various geometric and resolution biases tend to drive asymmetries down. Therefore, while a low asymmetry measurement does not guarantee that a galaxy is truly symmetric, when the background correction is properly applied a high asymmetry measurement does guarantee that the galaxy is indeed asymmetric.  But, for larger surveys, the asymmetry statistic can be used to select interesting galaxies as well as probe for differences in various populations (e.g. mergers versus non mergers or groups/clusters versus field galaxies).}

In addition to introducing the 3D asymmetry, we have also developed the `squared difference' asymmetry.  This asymmetry formulation allows for a \textcolor{black}{more} straightforward calculation of the contribution of noise to the measured asymmetry than the standard `absolute difference' asymmetry.  Unlike the absolute asymmetry, the background corrected squared difference asymmetry remains unbiased down to very low $S/N$.  This removes the need for some of the noise corrections developed in \citet{Giese2016} and \citet{Thorp2021} for low $S/N$ observations.

Based on these results, we expect that the 3D asymmetry for  WALLABY detections \textcolor{black}{with $D_{\hi}\ge 3$ beams} that have reliable masks can be calculated reliably.  This opens up many exciting avenues for future explorations, including the effects of environment on asymmetry, the connection between asymmetries and physical processes, and the use of asymmetries as a diagnostic for kinematic modelling.

\section*{acknowledgement}
\textcolor{black}{The authors thank the anonymous referee for their excellent suggestions.}
\textcolor{black}{KS acknowledges support from the Natural Sciences and Engineering Research Council of Canada (NSERC). \textcolor{black}{MG is supported by the Australian Government through the Australian Research Council's Discovery Projects funding scheme (DP210102103).}}

\section*{Data availability}

All mock cubes and analysis codes are available upon request to the corresponding author.  Additionally, the modified version of \textsc{MCGSuite} is available upon request.




\bibliographystyle{mnras}
\bibliography{Asymmetries.bib} 


\section*{Appendix}
\subsection*{3D Asymmetries Implementation}\label{Sec:Implementation}
\textcolor{black}{For this work, we have developed a software package called \textsc{3DACS} that calculates the 1D, 2D, and 3D asymmetries for a cubelet and mask.  The package is written in \textsc{Fortran} and consists of two distinct programs.  One calculates the asymmetry about a specific point, while the other attempts to find the point that minimizes the asymmetry.  }

\textcolor{black}{
Both programs can calculate either the `absolute difference' or `squared difference' asymmetry and can correct the measurements for the noise in the cubelet.  However, the background corrections given in Eqs. \ref{Eq:AbsBackground} and \ref{Eq:SqDBackground} depend on a uniform Gaussian noise.  Since the moment 0 map and velocity profile are calculated from the masked cube, this is not necessarily true.  To deal with this issue, the code approximates the noise for the 2D and 1D calculations based on the noise in the cube and the number of contributing cells to each pixel/channel (for 2D and 1D respectively).  Thus, the background corrected asymmetries for the 2D and 1D cases are only approximations.
}
\textcolor{black}{
As noted previously, it is important to utilize a symmetric mask about the rotation point.  For the specified center code, the 3D mask is first symmetrized and then the moment 0 map and velocity profile are calculated using this mask.  This means that the 2D map and 1D profile have an effectively symmetric mask.  The situation is more complicated when attempting to minimize the asymmetry as the cube, map, and profile are minimized independently.  While the mask can be resymmetrized at each trial center for the 3D cube, this is impossible for the map and profile as those are derived from the masked 3D cube.  Instead, for the minimization code, the map and profile are calculated using the original mask and are not adjusted/resymmetrized during the asymmetry minimzation steps.  This is an unfortunate limitation of the minimization procedure.  It should also be noted here that the asymmetry minimization code is significantly slower than using a defined center of rotation as it requires 1-2 orders of magnitude more pair calculations.
}

\bsp	
\label{lastpage}
\end{document}